\documentclass[11pt,a4paper]{article}

\usepackage[a4paper,margin=1in]{geometry}
\usepackage[T1]{fontenc}
\usepackage[utf8]{inputenc}
\usepackage{lmodern}
\usepackage{amsmath,amssymb}
\usepackage{graphicx}
\usepackage{xcolor}
\usepackage{cite}
\usepackage{booktabs}
\usepackage{url}
\usepackage{microtype}
\usepackage[hidelinks]{hyperref}

\title{Hollow-Core Fiber for Long-Span Optical Frequency Transfer: Improved Instability and Extended Single-Span Reach}

\author{%
\begin{minipage}{0.95\textwidth}\centering
Qian Zhou$^{1,2,\dagger}$, Ru Yuan$^{1,2,\dagger}$, Xiang Zhang$^{1,2,*}$, Yu Hua$^{1,2}$, Huibo Hong$^{1,2}$,\newline
Bo Liu$^{1,2}$, Rongduo Lu$^{3}$, Dawei Ge$^{3}$, Liuyan Han$^{3}$, Yucan Zhang$^{1,2}$,\newline
Yiting Liu$^{1,2}$, Dan Wang$^{1,2}$, Ruifang Dong$^{1,2,*}$, Tao Liu$^{1,2,*}$, and Shougang Zhang$^{1,2}$\\[0.75em]
$^{1}$National Time Service Center, Chinese Academy of Sciences, Xi'an 710600, China\\
$^{2}$Key Laboratory of Time and Frequency Standards, Chinese Academy of Sciences, Xi'an 710600, China\\
$^{3}$China Mobile Research Institute, Beijing 100053, China\\[0.75em]
$^{\dagger}$These authors contributed equally to this work.\\
$^{*}$Corresponding authors: Xiang Zhang (zhangxiang@ntsc.ac.cn), Ruifang Dong (dongruifang@ntsc.ac.cn), and Tao Liu (taoliu@ntsc.ac.cn).
\end{minipage}
}

\date{\today}

\begin{document}

\maketitle

\begin{abstract}

Phase-coherent optical frequency transfer is essential for optical clock networking, relativistic geodesy, and distributed precision metrology. However, realizing coherent optical networks spanning thousands of kilometers in standard single-mode fiber (SMF) generally requires densely distributed amplifiers or repeater stations together with complex operational control, while long-term instability remains limited by thermally driven residual phase fluctuations. Here we show that hollow-core fiber (HCF) can simultaneously improve transfer instability and relax the reach limitation of long-span optical frequency transfer. Compared with SMF, HCF exhibits lower fiber induced phase noise and shorter propagation delay, supporting improved short-term instability, while its much lower thermal sensitivity supports nearly one order of magnitude better long-term instability. In addition, for long-haul HCF links, no observable stimulated Brillouin scattering induced saturation is found up to the maximum available injected power of 34 dBm, whereas the threshold of an equal length SMF link remains only a few dBm. Together with the lower attenuation achievable in modern HCF, this enables ultra long single-span optical frequency transfer. Using a 152 km HCF link with an average attenuation of 0.18 dB/km, we demonstrate single-span optical frequency transfer, achieving a fractional frequency instability of $7.3 \times 10^{-21}$ at 10,000 s and a fractional uncertainty of $1.8 \times 10^{-20}$. These results establish HCF as a transmission medium that simultaneously improves instability and extends single-span reach, opening a practical route toward future intercontinental optical frequency networks with ultrahigh precision.

\end{abstract}



\section{Introduction}
State-of-the-art optical clocks have now achieved fractional instabilities and uncertainties at the $10^{-19}$ level\cite{Aeppli2024PRL,Marshall2025PRL,Liu2025PRL,McGrew2018Nature,Kim2023PRL}, opening the way to a new generation of precision measurement networks. By interconnecting optical clocks over long distances, such networks can enable remote clock comparisons at the highest level, relativistic geodesy, tests of fundamental physics, and other forms of distributed precision metrology\cite{Lindvall2025Optica,Grotti2024PRAppl,Lisdat2016NatCommun,Caldwell2025AOP,Formichella2024Optica}. Realizing this vision, however, requires transfer links whose added instability and uncertainty remain below the performance of the clock signals themselves.

Among the available techniques, optical fiber links remain the most practical route for phase-coherent optical frequency transfer over long distances. Conventional microwave satellite based methods generally provide performance only at the $10^{-16}$ level and are therefore insufficient for optical clock networking\cite{Bauch2015CRAS,Riedel2020Metrologia}. Free-space optical links have recently demonstrated remarkable performance over distances approaching a few hundred kilometers, but their susceptibility to atmospheric interruption makes continuous all-weather operation difficult\cite{Caldwell2023Nature,Shen2022Nature}. By contrast, optical fiber links have already enabled phase-coherent optical frequency transfer over thousand-kilometer distances with instabilities and uncertainties at the $10^{-20}$ level, and are therefore particularly attractive for building precision clock networks\cite{Droste2013PRL,Predehl2012Science,Chiodo2015OE,Raupach2015PRA}.

Despite this progress, realizing coherent optical networks spanning thousands of kilometers in standard single-mode fiber (SMF) remains fundamentally challenging. In long-haul SMF systems, the low stimulated Brillouin scattering (SBS) threshold severely restricts the maximum launch power, typically to only a few dBm~\cite{Kobyakov2010AOP,Terra2010OE}. Together with an attenuation of about 0.25 dB/km, this restricted power budget strongly constrains the achievable single-span distance and generally forces the long-haul SMF link to be segmented into relatively short spans bridged by bidirectional optical amplifiers or repeater laser stations. Existing thousand-kilometer SMF based optical frequency network illustrate this tradeoff well. Whether implemented as heavily amplified single span links or as cascaded national metrological networks using repeater laser stations, they rely on substantial intermediate infrastructure to overcome fiber loss and maintain phase coherent transfer\cite{Droste2013PRL, Chiodo2015OE, Lopez2012OE,Cantin2021NJP}. These intermediate nodes require additional phase locking, polarization control, and bidirectional stabilization, while simultaneously introducing excess noise, increasing control burden, and adding more potential failure points that raise the risk of cycle slips and link interruptions.

In addition, the limitations of SMF are not confined to transmission reach. Its relatively large propagation delay restricts the achievable bandwidth and noise-suppression efficiency of active phase noise cancellation, thereby constraining short-term transfer instability. On longer timescales, stabilized long-haul SMF links are often limited by residual phase fluctuations associated with polarization-related non-reciprocity and forward and backward optical frequency asymmetry, which remain correlated with the underlying thermally driven link noise and can limit the attainable long-term instability. This limitation becomes particularly significant in ultra-long-haul links and may ultimately constrain the transfer instability to the \(10^{-20}\) level\cite{Xu2025CPL,Chen2025arXiv2067km,Xu2021OptExpress_Nonreciprocity}. As a result, the performance of long-distance SMF-based optical frequency transfer is constrained not only by the fiber itself, but also by the complexity and operational fragility of densely cascaded architectures.

Hollow-core fiber (HCF), in which light propagates predominantly in an air core rather than silica, offers an attractive alternative transmission medium because it can simultaneously reduce light-matter interaction, propagation delay, and nonlinear coupling\cite{Slavik2015SciRep,Zhang2022LSA,Sakr2020NatCommun,Cooper2023Optica,Mulvad2022NatPhoton}. These properties are particularly relevant to optical frequency transfer. First, reduced light-matter interaction is expected to weaken the coupling of environmental perturbations to the optical phase, while the lower effective refractive index shortens the propagation delay and relaxes the delay-limited constraint of phase noise cancellation\cite{Feng2023OL,Fokoua2023AOP}. In addition, the much lower thermal sensitivity of HCF is expected to suppress temperature-induced phase drift and thereby provide favorable conditions for improved long-term instability. Taken together, these features suggest that HCF may support improved transfer instability on both short and long timescales.

A second advantage of HCF is its potential to substantially extend the achievable single-span transmission distance. Because nonlinear interaction is strongly suppressed, HCF can support much higher injected power than SMF before reaching the SBS threshold. At the same time, recent progress in HCF fabrication has reduced attenuation at 1550 nm to as low as 0.05 dB/km\cite{Ding2025ECOC,Petrovich2025NatPhoton,Ge2026JLT}, below the long-standing benchmark of conventional SMF~\cite{Khrapko2024IPTL}. The combination of much higher allowable launch power and ultralow loss provides a clear route toward optical frequency transfer over much longer single spans than are accessible in conventional SMF systems. Such an increase in span is important not only for reducing infrastructure requirements, but also for mitigating excess noise, operational complexity, and loss-of-lock risks associated with intermediate repeater lasers and polarization-control actions.

The potential of HCF has already been highlighted in related applications such as optical frequency comb dissemination, where it has shown clear advantages over SMF\cite{Feng2022LPR}. However, these results do not yet establish whether such advantages can be converted into meaningful system level gains for long haul phase-coherent optical frequency transfer. In particular, it remains unclear to what extent HCF can simultaneously alleviate the main instability limitations of the link and relax the distance constraint that arises from the restricted power budget of conventional SMF systems.

In this work, we experimentally investigate HCF for long-span optical frequency transfer by directly comparing a 152 km HCF link with an equal length SMF link. We address the problem from two complementary aspects. First, we measure the thermal response and transfer instability of HCF and SMF, and show that HCF exhibits reduced thermal sensitivity, smaller temperature related non-reciprocal noise, lower fiber-induced noise, and reduced propagation delay, which together are favorable for improving both short term and long term stability. Second, we study the high power transmission behavior, SBS threshold, beat note SNR, and power budget of the two fibers, and show that HCF supports much higher launch power and a substantially larger allowable loss budget than SMF. To the best of our knowledge, this work constitutes the first optical frequency transfer experiment over a hundred kilometer scale HCF link. The 152 km HCF system achieves a transfer instability of \(7.3\times10^{-21}\) at an integration time of 10,000 s and reaches an uncertainty at the \(10^{-20}\) level. These results provide a direct basis for assessing the potential of HCF for future ultra-long single-span optical frequency transfer and future long-haul optical clock networks.

\begin{figure}[!htbp]
  \includegraphics[width=\linewidth]{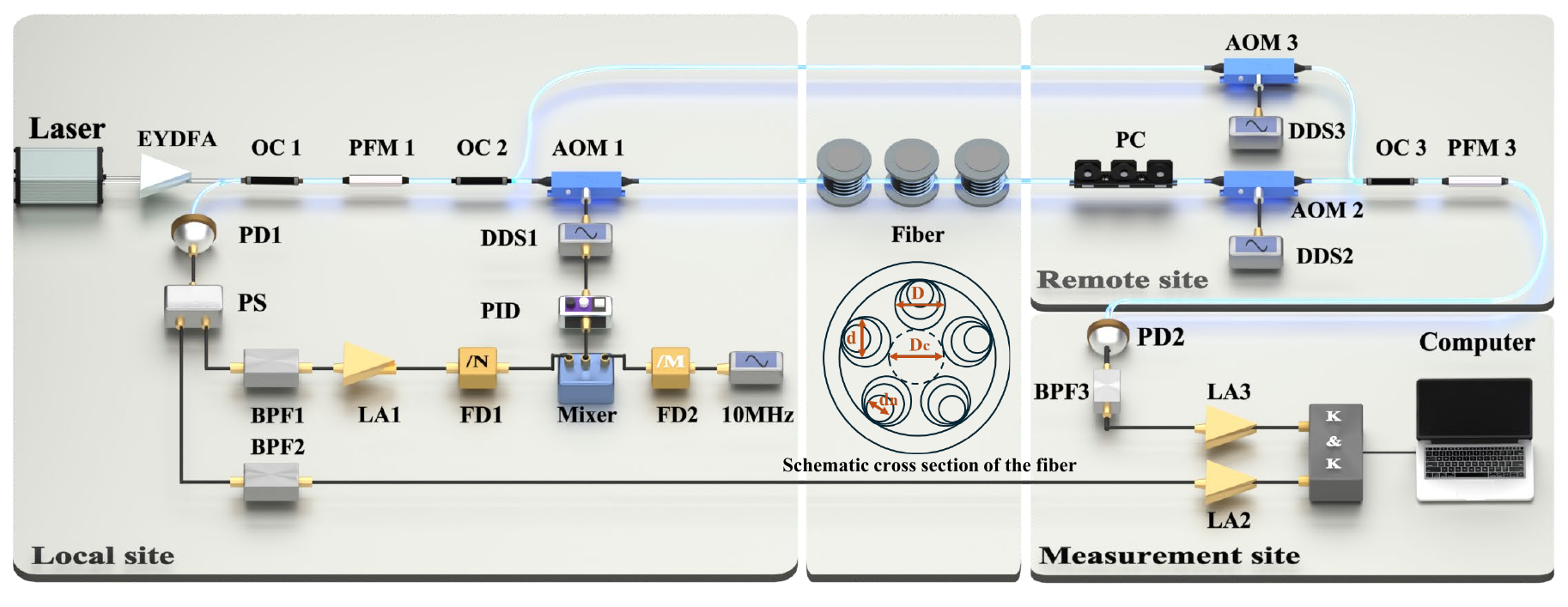}
  \caption{Experimental setup for optical frequency transfer over HCF and SMF links. The inset shows a schematic cross section of the HCF. 
EYDFA, erbium--ytterbium-doped fiber amplifier; OC, optical coupler; PFM, partial Faraday mirror; AOM, acousto-optic modulator; DDS, direct digital synthesizer; PD, photodetector; PS, power splitter; BPF, band-pass filter; LA, low-noise amplifier; FD, frequency divider; PID, proportional--integral--derivative controller; PC, polarization controller; and K\&K, frequency counter. }
  \label{fig:boat1}
\end{figure}
\section{Experiment setup}
The experimental setup is shown in Fig.~1. A commercial narrow-linewidth fiber laser operating at \(1550.12~\mathrm{nm}\), with a linewidth of approximately \(100~\mathrm{Hz}\) and an output power of \(10~\mathrm{dBm}\), served as the optical source (E15, NKT Photonics). The optical signal was first amplified by an erbium-ytterbium-doped fiber amplifier (EYDFA) and then directed to a partial Faraday mirror (PFM1) through optical coupler (OC1). PFM1 separated the incident light into two components: the reflected light served as the local reference for link phase-noise compensation, while the transmitted light propagated toward the transfer link.The signal transmitted through PFM1 was subsequently split by a \(90/10\) optical coupler (OC2). The \(10\%\) branch was used as the out-of-loop reference signal and frequency shifted by AOM3 driven at \(33~\mathrm{MHz}\), whereas the remaining \(90\%\) branch passed through AOM1 and was launched into the fiber link. AOM1 was operated at \(110~\mathrm{MHz}\), providing both the frequency shift for the propagating signal and the actuator for active compensation of link induced phase noise. At the remote site, the signal emerging from the fiber link was frequency shifted by AOM2 operated at \(50~\mathrm{MHz}\). A portion of the signal was tapped for transfer-performance characterization, while the remaining part was reflected by PFM2 and propagated back to the local end through the same fiber link. At the local end, the returned signal was heterodyned with the local reference at photodetector PD1, producing a beat note that contains the round-trip link phase noise. This beat note was filtered by a band pass filter centered at \(320~\mathrm{MHz}\), processed by a phase-locked loop, and fed back to AOM1, thereby enabling active compensation of the link-induced phase noise. In addition, for the out-of-loop characterization, the out-of-loop reference light frequency-shifted by AOM3 was heterodyned with the transmitted signal at PD2, generating a \(127~\mathrm{MHz}\) beat note for evaluating the transfer performance of the link. To eliminated the phase drift introduced by the out-of-loop reference arm, a \(66~\mathrm{MHz}\) reference beat note was simultaneously detected at PD1 at the local end. This signal contains twice the phase noise of the out-of-loop reference arm and can therefore be used in post-processing to subtract the reference arm noise contribution from the out-of-loop measurement, enabling a more accurate evaluation of the link transfer performance\cite{Qiu2025TIM}.The post processed 160~MHz frequency data were therefore used to characterize the link transfer performance, with the 127~MHz and 66~MHz frequency data both recorded using a K\&K frequency counter operated in the $\Lambda$-type counting mode with a gate time of 1~s.

The HCF employed in this work was a DNANF-5 HCF, where DNANF stands for double layer nested anti-resonant nodeless fiber. As shown in the inset of Fig.~1, the fiber has a hollow core diameter, $D_{\mathrm{c}}$, of approximately 32~$\mu$m. The surrounding capillary tubes form the anti-resonant cladding, which confines light through the anti-resonant reflection, while the nested inner capillaries further reduce confinement loss and suppress parasitic mode coupling\cite{Sakr2020NatCommun,Poletti2014OE}. The characteristic diameters of the anti-resonant elements, denoted by $D$, $d$, and $d_{\mathrm{n}}$, are approximately 30~$\mu$m, 22.8~$\mu$m, and 7.9~$\mu$m, respectively. The core is filled with nitrogen at a pressure of approximately 0.2 atm. The experimental link has a total length of 152 km and exhibits an average attenuation of approximately 0.18 dB/km at 1550 nm, including both splice losses and connector insertion losses.

\section{Improved frequency transfer instability in HCF}
The distinctive properties of HCF, especially its shorter propagation delay and reduced thermal sensitivity, suggest that it may improve both the short-term and long-term instability of optical frequency transfer. To test this expectation, we first compare the thermal response of HCF and SMF experimentally, then evaluate their transfer instability under free-running and stabilized conditions, and finally analyze the long-distance behavior through simulations of a 1000 km buried fiber link.

\subsection{Reduced propagation delay and phase noise for improved short-term instability}

Short-term frequency transfer instability is jointly determined by the fiber induced phase noise of the link and the delay limited performance of active phase noise cancellation. Because HCF combines weaker glass-mediated phase noise coupling with a lower effective refractive index, it is expected to outperform SMF in both respects. To quantify this advantage, we measured and compared the phase noise power spectral densities (PSDs) of a 152 km HCF link with that of an equal length SMF link. The phase noise were measured using a Microchip 53100A analyzer with a resolution bandwidth of 5 Hz over an offset-frequency range from 1 Hz to 10 kHz.

\begin{figure}[!htbp]
  \centering
  \includegraphics[width=0.7\linewidth]{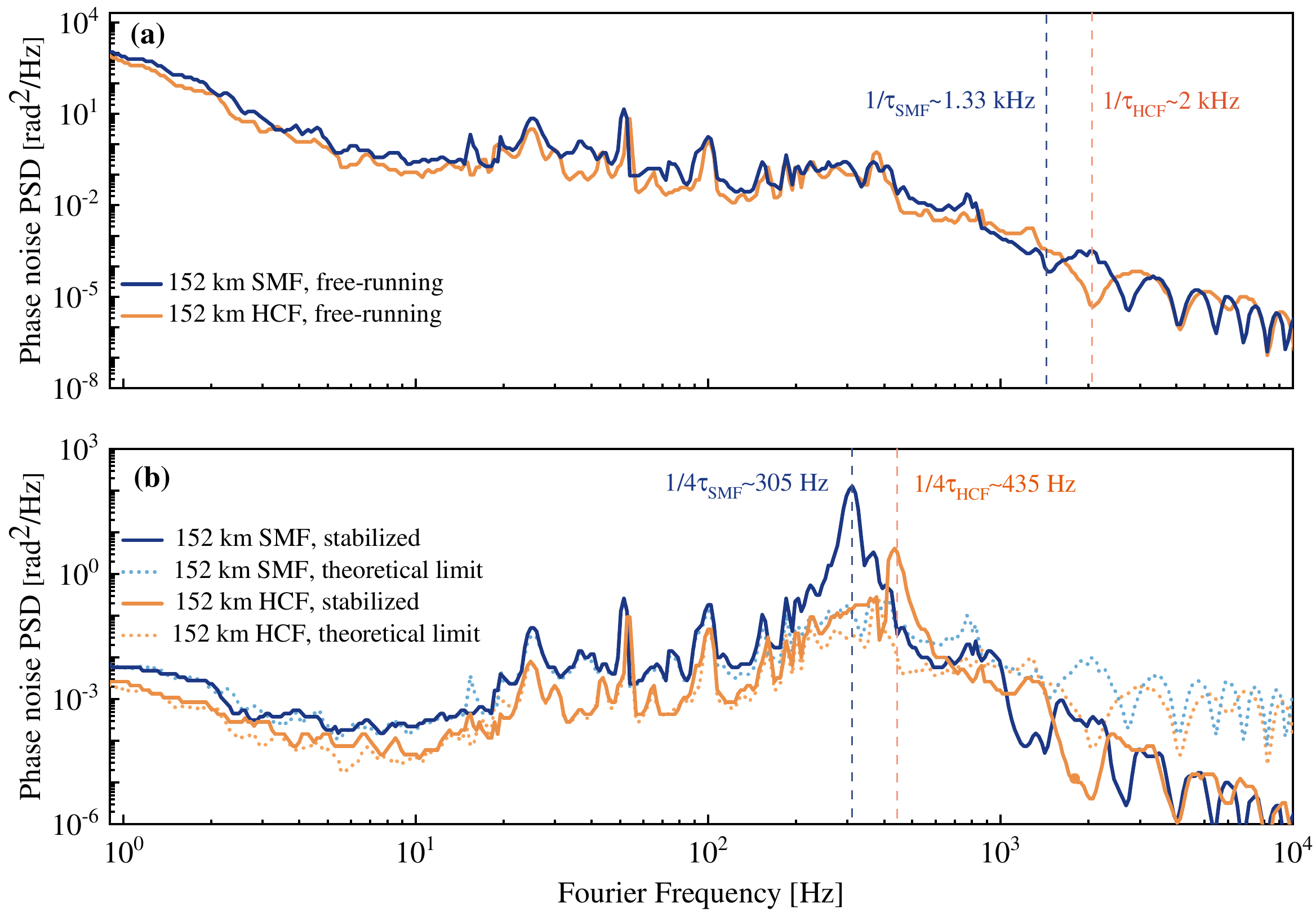}
  \caption{Phase noise PSDs of the 152~km SMF and HCF links. 
\textbf{a)} Free-running phase noise PSDs of the SMF and HCF links, shown as blue and orange curves, respectively. 
The dashed vertical lines indicate the characteristic inverse propagation delays of the two links,\(1/\tau_{\mathrm{SMF}}\sim1.33~\mathrm{kHz}\)and \(1/\tau_{\mathrm{HCF}}\sim2~\mathrm{kHz}\). \textbf{b)} Stabilized phase noise PSDs of the SMF and HCF links, together with the corresponding theoretical delay limited noise suppression curves.  The SMF and HCF results are shown in blue and orange, respectively. The dashed vertical lines indicate the approximate delay-limited servo bandwidths of the SMF and HCF links. }
  \label{fig:boat2}
\end{figure}

Figure 2(a) shows the measured phase noise PSDs of the two 152 km fiber links under free-running conditions. Under the same experimental environment, the HCF and SMF links exhibit similar overall spectral features over the range from 1 Hz to 10 kHz, indicating that the dominant noise processes are of the same general origin. In particular, both spectra show an approximately $f^{-4}$ dependence over 1 Hz-10 Hz and $f^{-1}$ dependence over 10 Hz-1 kHz, which are broadly consistent with low frequency drift processes and flicker type phase noise. While the spectral shape is similar, the corresponding noise level remains consistently lower in the HCF link. This reduction can be attributed primarily to the much smaller overlap of the optical mode in HCF with the surrounding glass, which weakens glass-mediated acoustic and optomechanical coupling and reduces the sensitivity of the link to environmentally induced phase perturbations\cite{Slavik2015SciRep,Iyer2020APLPhotonics}. Integrating the phase noise PSD over the analyzed Fourier-frequency range yields an integrated phase variance of 312 rad$^{2}$ for the 152 km HCF link and 536 rad$^{2}$ for the equal length SMF link, corresponding to a reduction by a factor of about 1.7 in HCF. This implies an approximately $\sqrt{1.7}\approx1.3$  fold improvement in short-term frequency instability under free running conditions.

In addition to lower fiber-induced phase noise, HCF is expected to exhibit a shorter propagation delay because light is guided predominantly in air rather than silica. The propagation delay can also be inferred from the first spectral dip in the free-running phase noise spectrum. As shown in Fig. 2(a), the first dip of the 152 km HCF link occurs at approximately 2 kHz, whereas that of the 152 km SMF link is observed at approximately 1.33 kHz. According to the relation $f_{dip}\sim1/\tau = c/(nL)$, the dip frequency $f_{dip}$ is inversely proportional to the single-trip propagation delay $\tau$. These measured values therefore indicate that, the propagation delay of the HCF link is approximately 1.5 times shorter than that of the equal length SMF link.

Figure~2(b) compares the stabilized phase noise PSDs of the two links with their corresponding theoretical noise suppression limits, given by \(S_{\mathrm{res}}(f)\approx \frac{1}{3}(2\pi f\tau)^2 S_{\mathrm{free}}(f)\)\cite{Newbury2007OL}. According to this relation, a 1.5 fold reduction in propagation delay implies, in principle, a 1.5 fold increase in the achievable servo bandwidth and a 2.25 fold improvement in the delay limited noise suppression capability. The experimentally measured servo bandwidths are 305~Hz for the SMF link and 435~Hz for the HCF link, in good agreement with the theoretical values of 333~Hz and 450~Hz, respectively, predicted from the propagation delay according to \(1/(4\tau)\)\cite{Williams2008JOSAB}. In addition, the measured stabilized phase noise PSDs of both links agree well with their respective theoretical limits, confirming that both systems operate close to the delay limited suppression regime. Combined with the 1.7 fold reduction in the integrated free-running phase variance established above, the shorter propagation delay of HCF is expected to yield an overall reduction in the residual phase noise power by about \(1.7 \times 2.25 \approx 3.8\). This expectation is consistent with the measured result in Fig.~2(b), which shows that the residual phase noise PSD of the HCF link is lower than that of the equal length SMF link by about 3.6 times. Therefore, under identical link lengths and environmental conditions, HCF is expected to reduce the residual phase noise power by about a factor of 3.8, corresponding theoretically to an approximately 1.9 fold improvement in short-term frequency instability under stabilized conditions.
It should be emphasized that the short-term instability improvement factor discussed here is valid only when the stabilized residual noise remains dominated by the fiber-induced noise, and is therefore more applicable to HCF links spanning hundreds of kilometers or more. For kilometer scale links, where the intrinsic link noise is relatively low, the stabilized residual noise is more likely to be limited by the system noise floor. As a result, the experimentally observable short-term instability improvement of HCF relative to SMF is expected to be reduced.

\subsection{Lower thermal sensitivity for improved long-term instability}
In optical frequency transfer systems, long-term phase fluctuations induced by ambient temperature variations along the fiber link can be expressed as,
\begin{equation}
\delta \phi_{\mathrm{free}}(t)=\omega_{0}\gamma L_{\mathrm{link}}\,\delta T_{\mathrm{link}}(t),
\end{equation}
Here, $\omega_{0}$ is the angular frequency of transmitted light, and $\gamma$ denotes the thermal sensitivity coefficient, defined as the induced time error per unit fiber length for a unit temperature change. In conventional solid core SMF, $\gamma$ is mainly determined by the thermo-optic contribution, whose magnitude is typically on the order of $10^{-5}\,\mathrm{K^{-1}}$, whereas the thermal expansion contribution is only on the order of $10^{-7}\,\mathrm{K^{-1}}$. By contrast, in HCF, light is guided predominantly in the air core, so the effective thermo-optic contribution is strongly suppressed. As a result, under the same temperature variation, an equal length HCF link is expected to exhibit much smaller thermally induced phase drift than an SMF link.

Recent studies have shown that, although most reciprocal thermal phase fluctuations can be efficiently canceled, residual non-reciprocal noise remains a fundamental limitation to the long-term instability of coherent optical frequency transfer, especially for long links. This residual phase fluctuation can be written as\cite{Chen2025arXiv2067km,Xu2021OptExpress_Nonreciprocity,Xu2019OptExpress_Reciprocity}
\begin{equation}
\delta \phi_{\mathrm{res}}(t)
=(\varepsilon B_{xy}+\frac{\Delta \omega}{\omega_{0}})\delta \phi_{\mathrm{free}}(t)
=\beta L_{\mathrm{link}}\,\delta T_{\mathrm{link}}(t),
\label{eq:residual_phase}
\end{equation}
Here, the first term represents the polarization related non-reciprocal contribution, where $B_{xy}$ denotes the fiber birefringence coefficient and $\varepsilon$ is temperature dependent averaging factor ($\left|\varepsilon\right|\leq1$). The second term represents the non-reciprocal phase contribution caused by the forward-backward optical frequency asymmetry introduced mainly by the remote site AOM2 ($\Delta \omega\sim2\times 50$ MHz). Since both contributions scale with link length and temperature fluctuation, they are conveniently grouped into an effective thermal non-reciprocity coefficient $\beta$, which is used below to characterize the overall thermal related non-reciprocal response of the stabilized link.

\begin{figure}[!h]
\centering
  \includegraphics[width=0.8\linewidth]{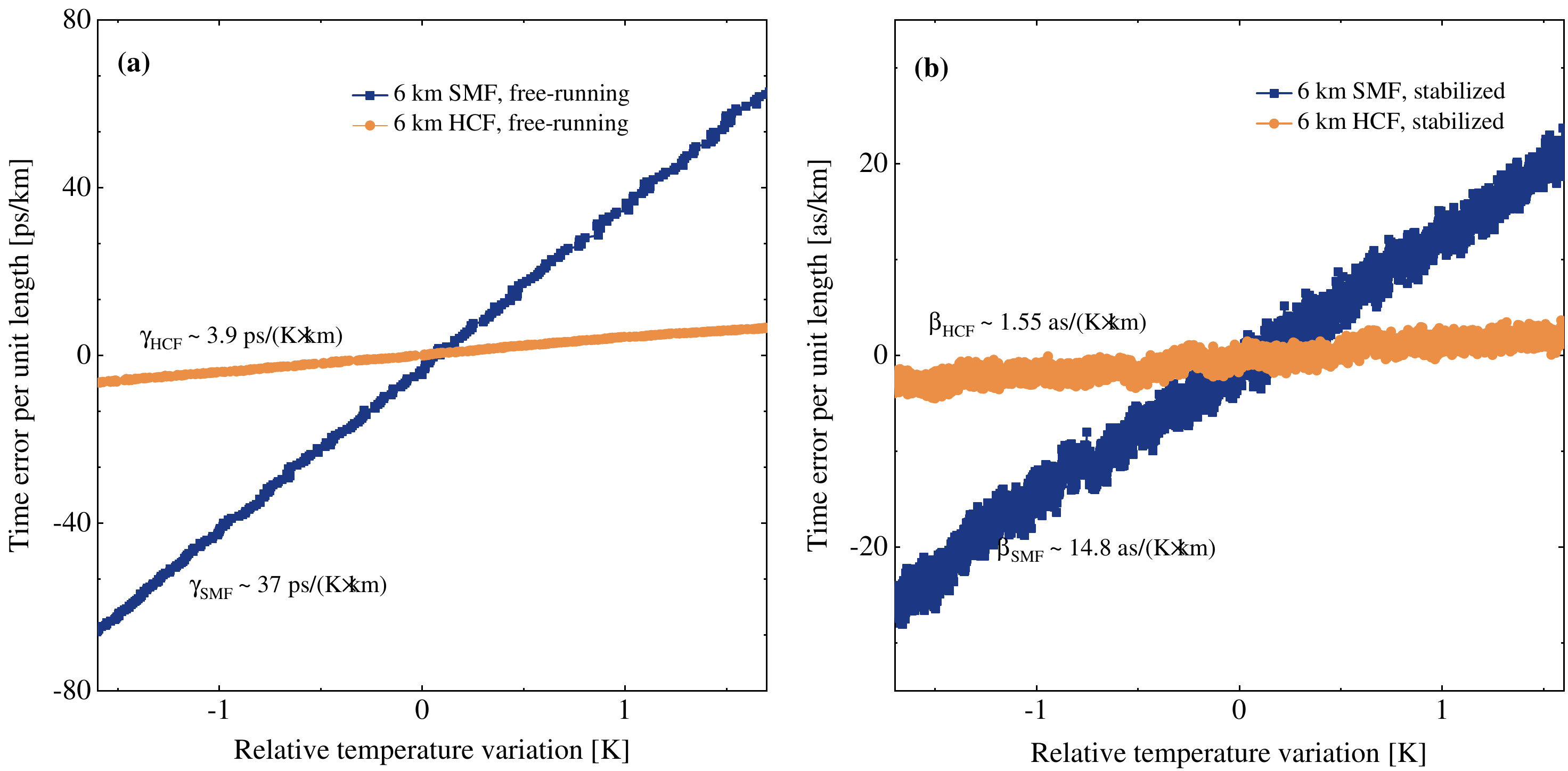}
  \caption{Thermal response coefficients of the 6~km SMF and HCF links. 
\textbf{a)} Free-running time error per unit length as a function of relative temperature variation for the SMF and HCF links, shown in blue and orange, respectively. 
Linear fits yield thermal sensitivities of \(\gamma_{\mathrm{SMF}}\sim37~\mathrm{ps/(K\cdot km)}\) and \(\gamma_{\mathrm{HCF}}\sim3.9~\mathrm{ps/(K\cdot km)}\). 
\textbf{b)} Stabilized residual time error per unit length as a function of relative temperature variation for the SMF and HCF links, shown in blue and orange, respectively. 
The extracted effective thermal non-reciprocity coefficients are \(\beta_{\mathrm{SMF}}\sim14.8~\mathrm{as/(K\cdot km)}\) and \(\beta_{\mathrm{HCF}}\sim1.55~\mathrm{as/(K\cdot km)}\). }
  \label{fig:boat3}
\end{figure}
To determine both the thermal sensitivity coefficient $\gamma$ and the effective thermal non-reciprocity coefficient $\beta$ for SMF and HCF, comparative optical frequency transfer experiments were carried out on fiber sections placed in a temperature-controlled chamber. Owing to the limited chamber size, 6 km fiber sections taken from the two 152 km links were used for comparison. During the temperature variation process, the chamber temperature was monitored in real time using a data logger (TC-08, Pico Technology), while the corresponding phase variations of the transmitted optical signal under the free-running and stabilized conditions were recorded using a K\&K frequency counter. For clarity, the measured phase variations are expressed here as equivalent time errors and normalized by the fiber length. The coefficients $\gamma$  and $\beta$ were then obtained from linear fits of the normalized time error to the measured temperature variation for the free-running and stabilized cases, respectively.

The results are summarized in Fig.~3, where Fig.~3(a) presents the free-running response and Fig.~3(b) the stabilized response. Linear fits to the data in Fig.~3(a) yield \(\gamma_{\mathrm{HCF}}\approx 3.9~\mathrm{ps/(K\!\cdot\!km)}\) and \(\gamma_{\mathrm{SMF}}\approx 37~\mathrm{ps/(K\!\cdot\!km)}\) are obtained. For a temperature excursion of 3.2 K, this corresponds to induced time errors of about 11.7 ps/km for HCF and 111 ps/km for SMF, indicating that the thermal sensitivity of HCF is lower by about one order of magnitude. From Fig.~3(b), the effective thermal non-reciprocity coefficients are extracted as \(\beta_{\mathrm{HCF}}\approx 1.5~\mathrm{as/(K\!\cdot\!km)}\) and \(\beta_{\mathrm{SMF}}\approx 14.8~\mathrm{as/(K\!\cdot\!km)}\), corresponding to induced time errors of about 4.9 $\mathrm{as/km}$ for HCF and 47.3 $\mathrm{as/km}$ for SMF under the same temperature excursion. In addition, the birefringence coefficients of the DNANF-5 HCF and the SMF are found to be comparable, both being on the order of $1\times10^{-7}$. Therefore, the much lower $\beta_{HCF}$ mainly originates from its substantially reduced thermal sensitivity rather than from a pronounced difference in birefringence. These results indicate that HCF can suppress both free-running thermal response and the residual thermal non-reciprocal response by about one order of magnitude, which theoretically supports an approximately 10 fold improvement in long-term frequency instability.

\begin{figure}[!h]
  \centering
  \includegraphics[width=0.55\linewidth]{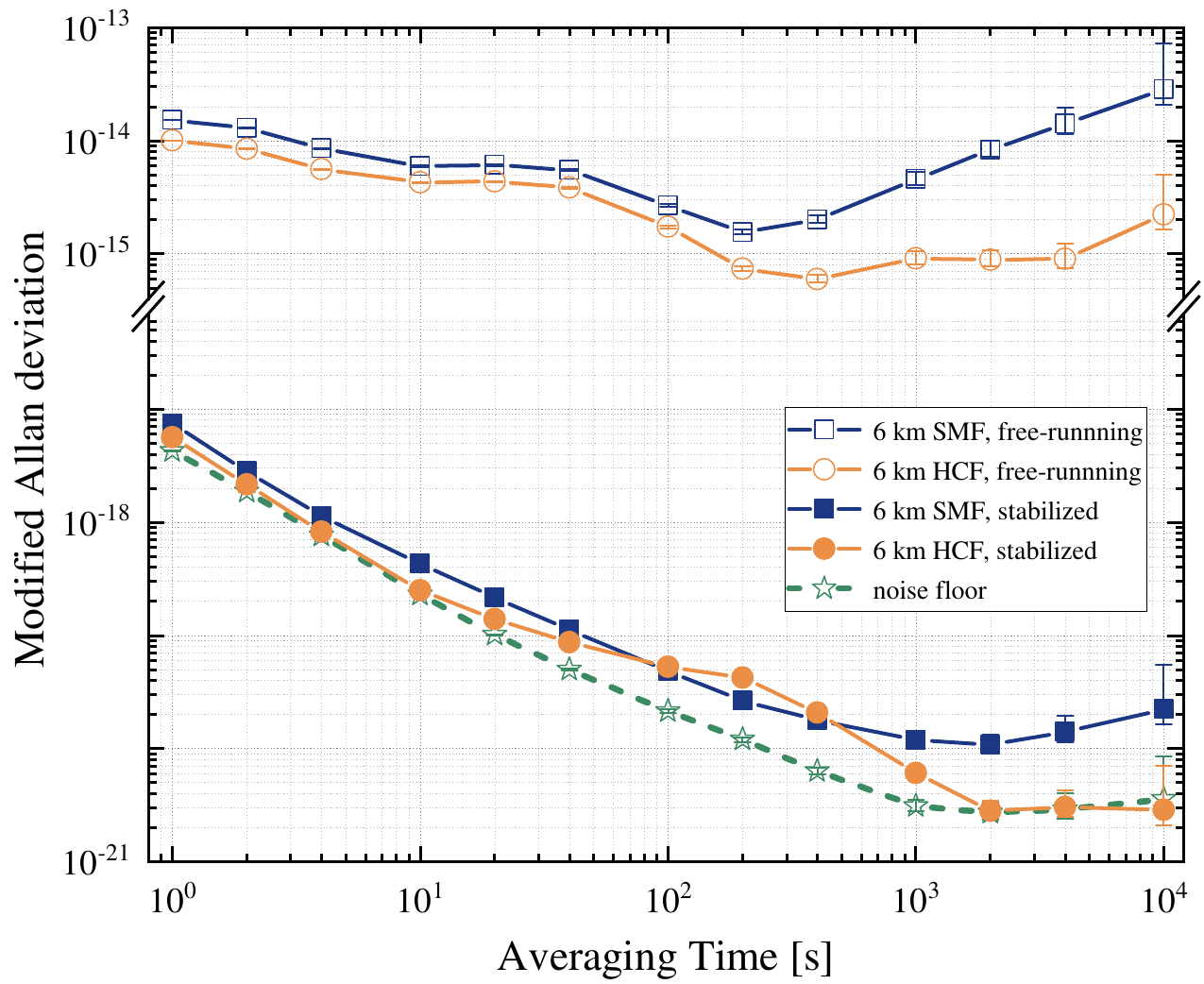}
  \caption{Fractional frequency instability comparison between 6 km SMF and HCF links. 
Modified Allan deviations of the 6~km SMF and HCF links under free-running and stabilized conditions. 
Open symbols show the free-running instability of the SMF and HCF links, shown in blue and orange, respectively. 
Filled symbols show the residual instability after active phase-noise cancellation. 
The green dashed curve denotes the system noise floor. }
  \label{fig:boat4}
\end{figure}

To further verify the impact of the measured thermal response on frequency transfer performance, we further compared the frequency transfer instability of 6 km HCF and SMF links in terms of the modified Allan deviation (MDEV). During a measurement interval of 43,000 s, the ambient temperature surrounding the fiber links drifted by approximately 13 K. As shown in Fig.~4, the 6 km free-running HCF link exhibited a frequency instability of \(1.0\times10^{-14}\) at 1 s averaging time, compared with \(1.5\times10^{-14}\) for the SMF link, corresponding to an improvement factor of about 1.5. As the averaging time increased, the instability of both links became increasingly dominated by ambient temperature fluctuations. At 10,000 s, the instability of the HCF link was \(2.2\times10^{-15}\), whereas that of the SMF link reached \(2.8\times10^{-14}\). Thus, under free-running conditions, the HCF link exhibited about one order of magnitude lower long-term instability than the SMF link, consistent with its substantially reduced thermal sensitivity. After active noise cancellation, the short-term instability of the HCF link remained lower than that of the SMF link, with values of \(5.6\times10^{-18}\) and \(7.4\times10^{-18}\) at 1 s, respectively. For a 6 km link, the difference in propagation delay between HCF and SMF is relatively small, so the corresponding difference in delay-limited noise suppression capability is also limited. The remaining short term advantage of HCF is therefore attributed mainly to its lower free-running noise. At longer averaging times, the fractional frequency instability of the HCF link continued to improve and reached \(2.9\times10^{-21}\) at 10,000 s, corresponding to the system noise floor, whereas the SMF link was limited to \(2.2\times10^{-20}\). Therefore, although the short-term improvement after stabilization is modest for kilometer scale links, the long-term instability of the HCF link remains nearly one order of magnitude better than that of the SMF link.

\begin{figure}[!ht]
\centering
\includegraphics[width=0.55\linewidth]{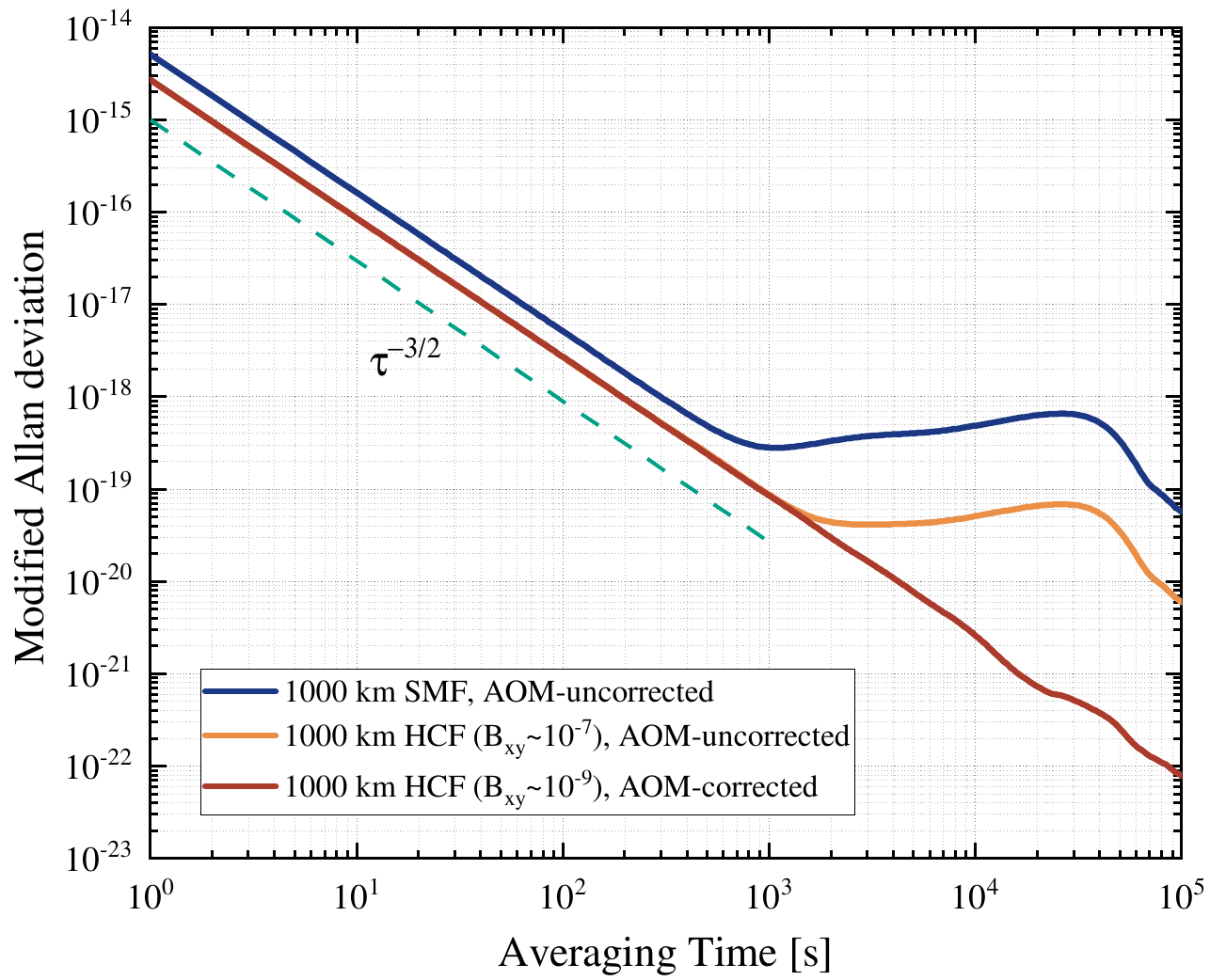}
\caption{Predicted fractional frequency instability of 1,000~km SMF and HCF optical-frequency transfer links. 
Modified Allan deviations are shown for an AOM-uncorrected SMF link, an AOM-uncorrected HCF link with \(B_{xy}\sim10^{-7}\), and an AOM-corrected HCF link with \(B_{xy}\sim10^{-9}\), represented by the blue, orange, and dark-red curves, respectively. 
The green dashed line indicates a \(\tau^{-2/3}\) reference slope.}
\label{fig:1000km_simulation}
\end{figure}

Using the thermal response parameters extracted above, we further simulated residual time error of a 1,000 km stabilized buried fiber link, with the main aim of comparing the long-term frequency transfer instability of SMF and HCF under realistic environmental perturbations. The residual time error of the stabilized fiber link was constructed from two contributions. The first was a short-term white phase noise background, whose amplitude here was adjusted to remain consistent with reported results~\cite{Droste2013PRL,Raupach2015PRA,Xu2025CPL}. The second was a temperature induced drift term derived from an eight day record of measured ambient temperature variation, assuming an effective temperature coupling factor of 0.2 for the buried fiber.

The simulation frequency transfer stabilities are summarized in Fig.~5. Daily temperature variation imposes a clear long-term limitation on the 1000 km SMF link, restricting the instability to the \(4.8\times10^{-19}\) level over averaging times from 1,000 s to 40,000 s, and to about \(7.5\times10^{-20}\) at one day. This behavior is consistent with reported long-haul optical frequency transfer performance and reflects the present limitation imposed by residual non-reciprocal noise in links at the thousand kilometer scale. Under otherwise identical conditions, the HCF link exhibits approximately one order of magnitude lower instability, reaching the \(4.8\times10^{-20}\) level over the same averaging time range and about \(7.6\times10^{-21}\) at one day. This improvement arises from the much smaller thermally induced non-reciprocal response of HCF, as established by the measurements above. Figure~5 also includes a more advanced HCF scenario. If an HCF with a birefringence coefficient at the \(10^{-9}\) level is employed\cite{Hou2026JLT}, and the frequency asymmetry contribution introduced by the remote AOM is accurately corrected, the instability contribution from the fiber link can be further reduced to the low \(10^{-23}\) level, corresponding to a further improvement of about three orders of magnitude relative to the SMF. Here, this value refers specifically to the contribution from the fiber link itself. The overall transfer instability would still depend on the noise floor of the transfer system, which may be further improved in future integrated implementations or frequency transfer architectures based entirely on HCF. These results indicate that HCF provides a realistic route toward suppressing the long-term non-reciprocal limitation of buried long-haul fiber links.

\section{Extended single-span capability enabled by HCF}
In phase-coherent optical frequency transfer, the maximum single-span distance is closely related to the by the useful signal power remaining after long-haul propagation, since this power sets the beat-note SNR available for phase-noise extraction and compensation. For a fixed detection configuration, the achievable span is therefore governed by the available launch power and the fiber attenuation. In conventional solid-core silica fibers, both factors are intrinsically constrained by the attenuation floor and the low SBS threshold~\cite{Kobyakov2010AOP,Sato2025JLT,Tamura2018JLT}. By contrast, HCF guides most of the optical field in air, thereby reducing light matter interaction and offering the potential for both substantially higher allowable launch power and lower attenuation. However, the long-haul transmission characteristics of narrow linewidth lasers in HCFs, which are essential for phase coherent optical frequency transfer, remain largely unexplored.

\subsection{Greatly increased SBS threshold and allowable launch power}

To investigate the response of SMF and HCF links under high launch powers, we constructed the experimental setup shown in Fig.~6. The high power transmission behavior were characterized by measuring the optical spectra of the forward and backward propagating signals after long-haul transmission, together with the beat note SNR. For the forward and backward propagating spectral measurements, OS2 was switched between Channels B and A, respectively, while OS1 was connected to Channel B for the backward propagating case. The optical spectra were recorded using an optical spectrum analyzer with a resolution of \(0.03~\mathrm{nm}\) (MS9740B, Anritsu). For the beat note SNR measurement, both optical switches were set to Channel A. The laser output was first divided into two arms by OC1. One arm was reflected by FM1 and used as the reference light, whereas the other arm was amplified by the EYDFA and launched into the fiber link through a circulator. After transmission through the fiber link, the signal was frequency-shifted by AOM1, reflected by FM2, and then propagated back along the same fiber link before exiting from the circulator. The returned signal was adjusted using a variable optical attenuator (VOA) and heterodyned with the reference light to generate a beat note, which was recorded using a radio-frequency spectrum analyzer (FSH4, Rohde \& Schwarz).
\begin{figure}[!h]
  \centering
  \includegraphics[width=\linewidth]{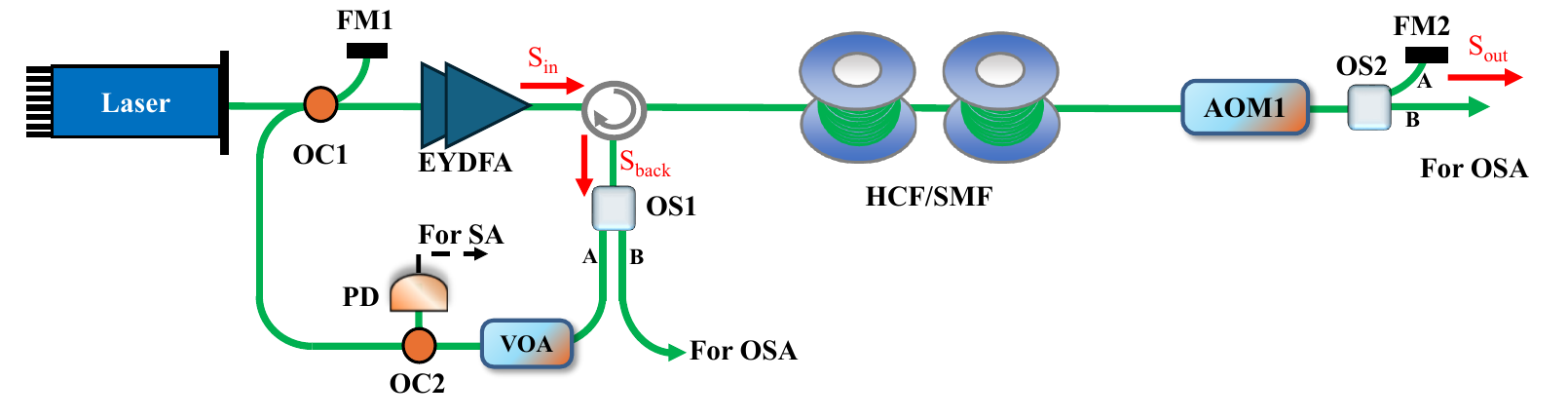}
  \caption{Experimental setup for high-power transmission characterization of HCF and SMF links. OC, optical coupler; FM, Faraday mirror; AOM, acousto-optic modulator; PD, photodetector; OS, optical switch; VOA, variable optical attenuator; SA, spectrum analyzer; OSA, optical spectrum analyzer. }
  \label{fig:boat6}
\end{figure}

Figure~7 compares the forward and backward propagating signals of the 152 km SMF and HCF links over a launch power range from 10 to 34~dBm. Figure~7(a) shows the forward propagating output signal power after transmission through the two links. For the HCF link, the output power increases monotonically with launch power, while the inferred link loss remains essentially unchanged, indicating the absence of observable power-dependent excess loss. By contrast, the output power of the SMF link saturates and is clamped near \(-30~\mathrm{dBm}\), indicating that it is already limited by SBS. The inset of Fig.~7(a) shows the corresponding output spectra at different launch powers. For the HCF link, the spectral envelope remains stable, exhibiting only an overall increase in amplitude. In contrast, the SMF spectrum shows clear saturation of the main signal peak, together with pronounced sidebands at an offset of approximately 0.08~nm whose intensity increases with launch power. To clarify the physical origin of this saturation behavior, we further monitored the backward-propagating spectra at Channel B of OS1, as shown in Fig.~7(b). For the HCF link, only an enhancement of the main backward signal peak is observed. In contrast, the SMF link exhibits a distinct backward-propagating Stokes wave at a wavelength offset of approximately \(0.08~\mathrm{nm}\) from the signal peak, with its intensity increasing markedly at higher launch powers. This offset is consistent with the Brillouin Stokes shift in silica fiber, confirming that a substantial fraction of the input optical power in the SMF link is converted into backward propagating Stokes components. As a result, the forward-propagating signal becomes power clamped, giving rise to the saturation observed in Fig.~7(a). In contrast, no observable SBS induced power saturation is found in the HCF link even at a launch power of \(34~\mathrm{dBm}\), indicating that its SBS threshold exceeds the maximum launch power available in our experiment.
\begin{figure}[!h]
  \centering
  \includegraphics[width=\linewidth]{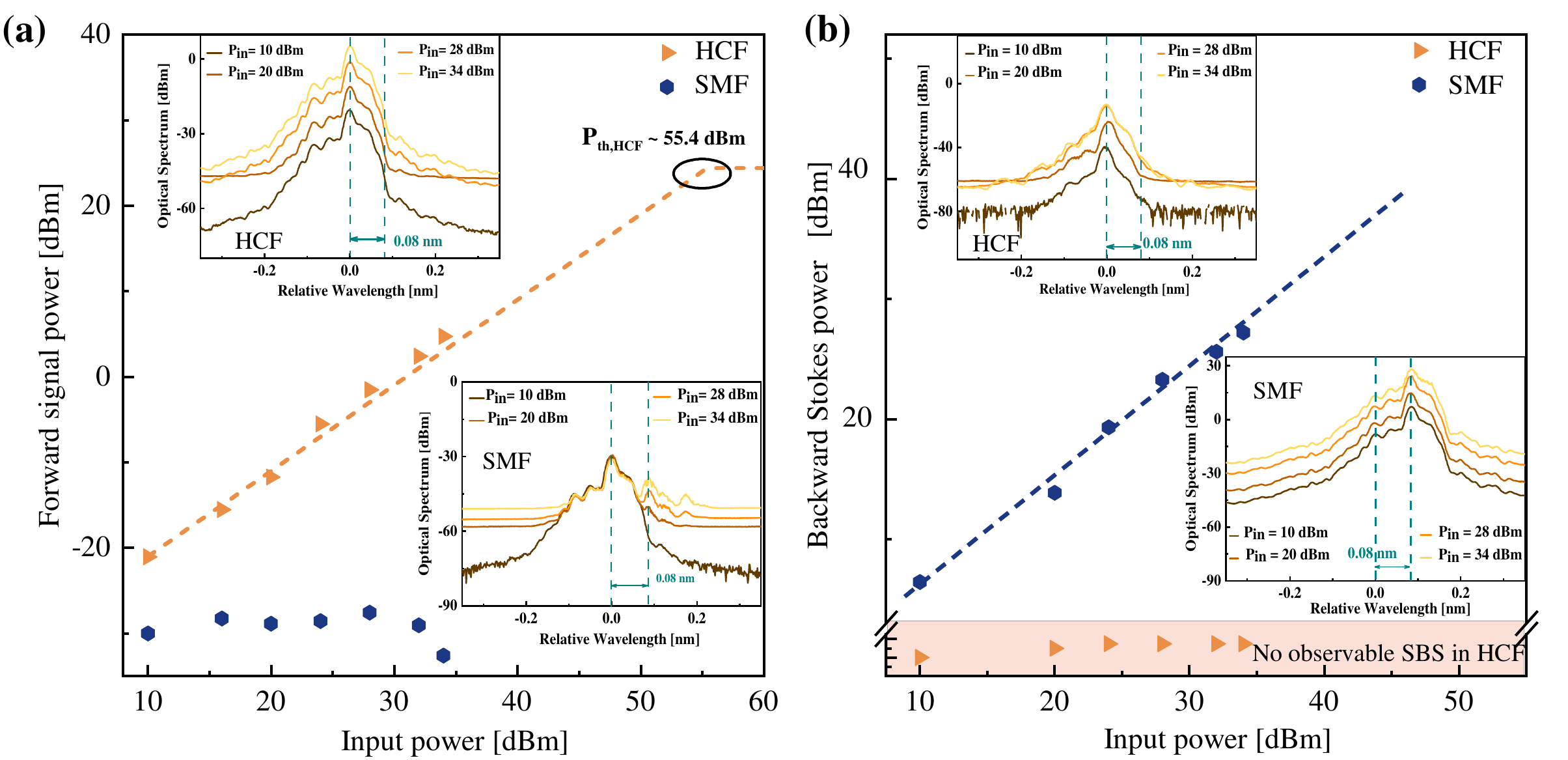}
  \caption{High-power transmission and SBS characteristics of the HCF and SMF links. 
\textbf{a)} Forward signal power as a function of input power for the HCF and SMF links, shown as orange and blue symbols, respectively. 
The HCF link shows a nearly linear power increase, whereas the SMF link exhibits clear power saturation. 
Insets show the forward output spectra of the HCF and SMF links at different input powers. 
\textbf{b)} Backward Stokes power as a function of input power for the HCF and SMF links, shown as orange and blue symbols, respectively. 
A rapid increase in backward Stokes power is observed in the SMF link, while no observable SBS is detected in the HCF link within the measured input-power range. 
Insets show the corresponding backward-propagating spectra of the HCF and SMF links. }
  \label{fig:boat7}
\end{figure}

Because the SBS threshold sets the maximum usable launch power and therefore limits the achievable single-span distance, a quantitative evaluation is required to assess the ultimate single-span capability of both HCF and SMF. To this end, we further analyze the SBS threshold using the theoretical model in Ref.~\cite{Kobyakov2010AOP},

\begin{equation}
P_{\mathrm{SBS,th}}=\frac{21S_{\mathrm{eff}}}{g_{\mathrm{B}}}\frac{\alpha}{1-e^{-\alpha L}},
\label{eq:sbs_threshold}
\end{equation}
where \(S_{\mathrm{eff}}\) is the effective mode area, \(g_{\mathrm{B}}\) is the Brillouin gain coefficient, \(\alpha\) is the linear power attenuation coefficient of the fiber, related to the commonly quoted attenuation in dB/km by \(\alpha=(\ln 10/10)\alpha_{\mathrm{dB}}\).

For the 152 km SMF link, using \(S_{\mathrm{eff,SMF}}\approx1\times10^{-10}~\mathrm{m^2}\), \(g_{\mathrm{B,SMF}}\approx5\times10^{-11}~\mathrm{m/W}\) and \(\alpha_{dB,SMF}\approx0.2~\mathrm{dB/km}\), the model yields an SBS threshold of approximately \(2.9~\mathrm{dBm}\), consistent with the experimentally observed few dBm limit for narrow-linewidth transmission in silica fiber. By contrast, for the 152 km HCF link used in this work, with \(S_{\mathrm{eff,HCF}}\approx8\times10^{-10}~\mathrm{m^2}\) and \(\alpha_{\mathrm{dB,HCF}}\approx0.18~\mathrm{dB/km}\). Since the hollow-core was filled with nitrogen at a static pressure of \(0.2~\mathrm{atm}\), the Brillouin gain coefficient is estimated to be \(g_{\mathrm{B,HCF}}\approx2\times10^{-15}~\mathrm{m/W}\) based on the model in Refs.~\cite{Yang2026LSA,Yang2020NatPhoton}. Substituting these values into Eq.~(\ref{eq:sbs_threshold}) gives an estimated SBS threshold of approximately \(55.4~\mathrm{dBm}\) for the 152 km HCF link. This value is more than 50 dB higher than that of the corresponding SMF link and far above the maximum launch power of \(34~\mathrm{dBm}\) available in our experiment, which explains the absence of observable SBS-induced power saturation in the HCF measurements.

In addition, in an optical frequency transfer link, the achievable single-span distance is determined not only by the launch power, but also by whether the returned optical power can maintain a sufficiently high beat-note SNR for stable phase detection and frequency counting. To quantify this constraint, we measured the beat-note SNR under different equivalent link-attenuation conditions at a fixed launch power of \(34~\mathrm{dBm}\). Here, the maximum permissible link loss is defined as the total equivalent attenuation for which the beat note SNR remains above the minimum level required for stable phase detection and frequency counting. To maximize the use of real fiber rather than relying entirely on VOA emulation, the measurement was performed in a loop back configuration using remote-end reflection. Accordingly, the horizontal axis in Fig.~8 represents the total equivalent attenuation experienced by the returned signal. For the HCF link, the one way attenuation of the 152 km span is about \(28~\mathrm{dB}\), corresponding to an equivalent starting attenuation of about \(62~\mathrm{dB}\) after round-trip propagation and 6 dB double-pass insertion loss comes from the remote AOM. The corresponding starting attenuation for the SMF link is about \(67~\mathrm{dB}\). Additional attenuation was then introduced with a VOA to emulate longer links. During the measurements, the spectrum analyzer was operated with a frequency span of \(1~\mathrm{MHz}\) and a resolution bandwidth (RBW) of \(30~\mathrm{kHz}\), while the reference light power was fixed at \(150~\mu\mathrm{W}\).

\begin{figure}[!h]
  \centering
  \includegraphics[width=0.5\linewidth]{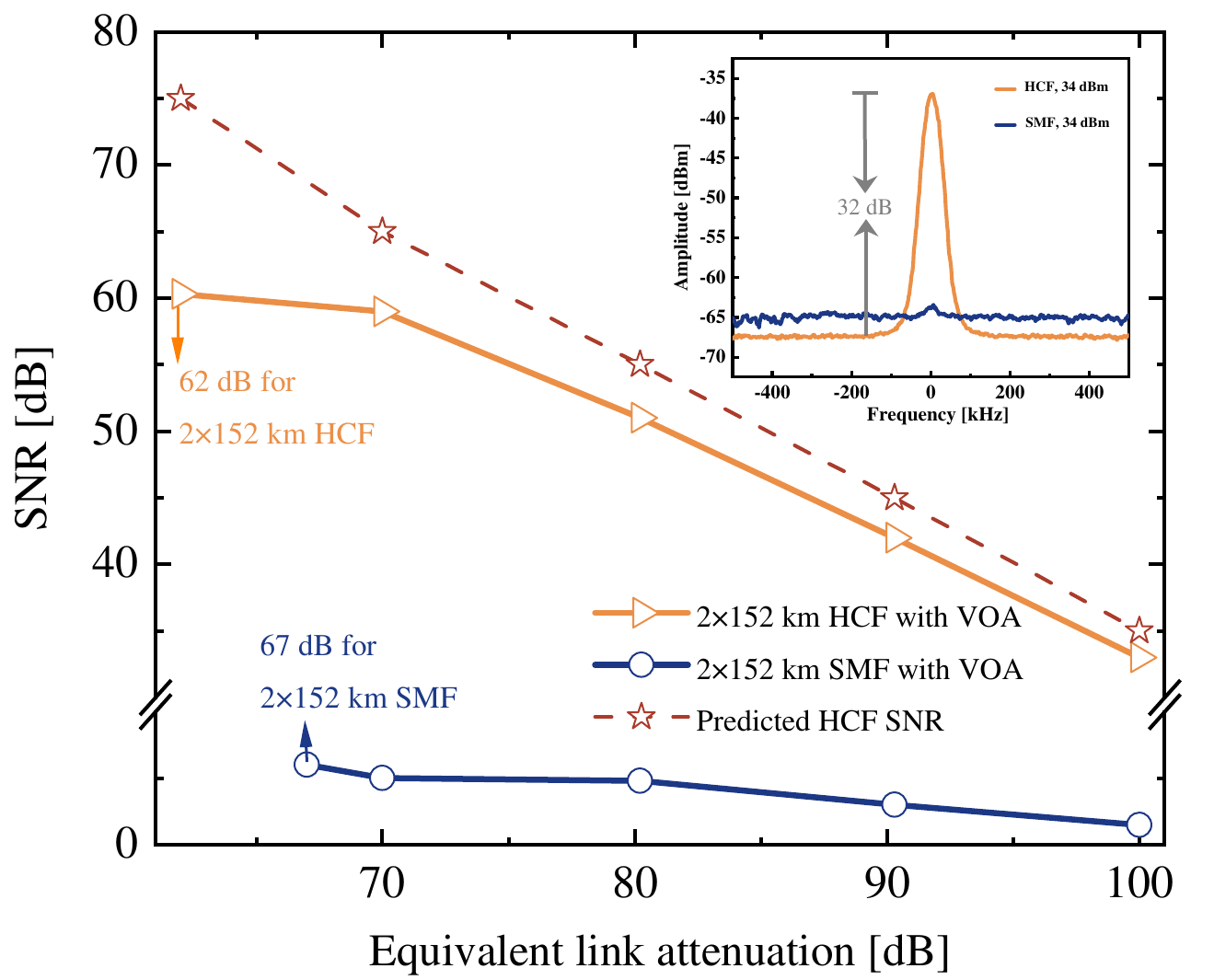}
  \caption{Beat note SNR under long distance transmission loss.The measured SNRs of the \(2\times152~\mathrm{km}\) HCF and SMF links are plotted as a function of equivalent link attenuation, with additional loss introduced by the VOA. The HCF and SMF results are shown as orange triangles and blue circles, respectively. The dashed curve shows the predicted HCF SNR. The arrows indicate the intrinsic losses of the HCF and SMF links. The inset shows the beat-note spectra of the HCF and SMF links at an link loss of \(100~\mathrm{dB}\).}
  \label{fig:boat8}
\end{figure}
As shown in Fig.~8, the beat note SNR after the HCF link decreases gradually with increasing equivalent link attenuation, yet still remains as high as \(32~\mathrm{dB}\) at an equivalent link attenuation of \(100~\mathrm{dB}\), leaving a substantial margin for stable phase detection and noise cancellation. By contrast, the SNR of the SMF link stays below \(6~\mathrm{dB}\) over the entire investigated attenuation range and approaches the noise floor at the highest attenuation. The inset further highlights this contrast, showing a clearly resolved beat note for the HCF link, whereas the SMF beat note is nearly buried in noise. At this attenuation, the measured HCF SNR agrees reasonably well with the theoretical value of \(35~\mathrm{dB}\)\cite{Teich1972JAP}. The remaining discrepancy is mainly attributed to stray-light interference in the optical path and nonlinear saturation in the integrated amplifier circuit of the photodetector. Together, these results demonstrate that, under high-power launch conditions, HCF provides a substantially larger allowable loss budget than SMF. They further indicate that the minimum received optical power required for stable system operation, denoted by \(P_{\mathrm{rec,min}}\), is on the order of \(-70~\mathrm{dBm}\) for the present receiver configuration\cite{Risaro2022PRAppl}. This system-level constraint provides an important basis for evaluating the ultimate single-span reach of phase-coherent optical frequency transfer.

\subsection{Power budget and potential for ultra-long signle-span transfer }
Based on the above analysis, the maximum single-span distance can be estimated using a simplified power-budget model,
\begin{equation}
L_{\max}\approx\frac{P_{th,SBS}-P_{\mathrm{rec,min}}}{\eta\,\alpha_{\mathrm{dB}}},
\label{eq:lmax}
\end{equation}

where the \(P_{\mathrm{th,SBS}}\) stands for the maximum usable launch power and \(P_{\mathrm{rec,min}}\) is on the order of \(-70~\mathrm{dBm}\) for the present frequency transfer configuration. Here, \(\eta\) denotes the path factor of the optical frequency-transfer scheme, with \(\eta=1\) for the two-way scheme and \(\eta=2\) for the round-trip scheme. In this simplified treatment, the additional fixed insertion loss of about \(6~\mathrm{dB}\) in the round-trip configuration is neglected. Therefore, under the same power-budget conditions, the maximum one-way span of the two-way scheme is approximately twice that of the round-trip scheme.

\begin{figure}[!h]
  \centering
  \includegraphics[width=0.6\linewidth]{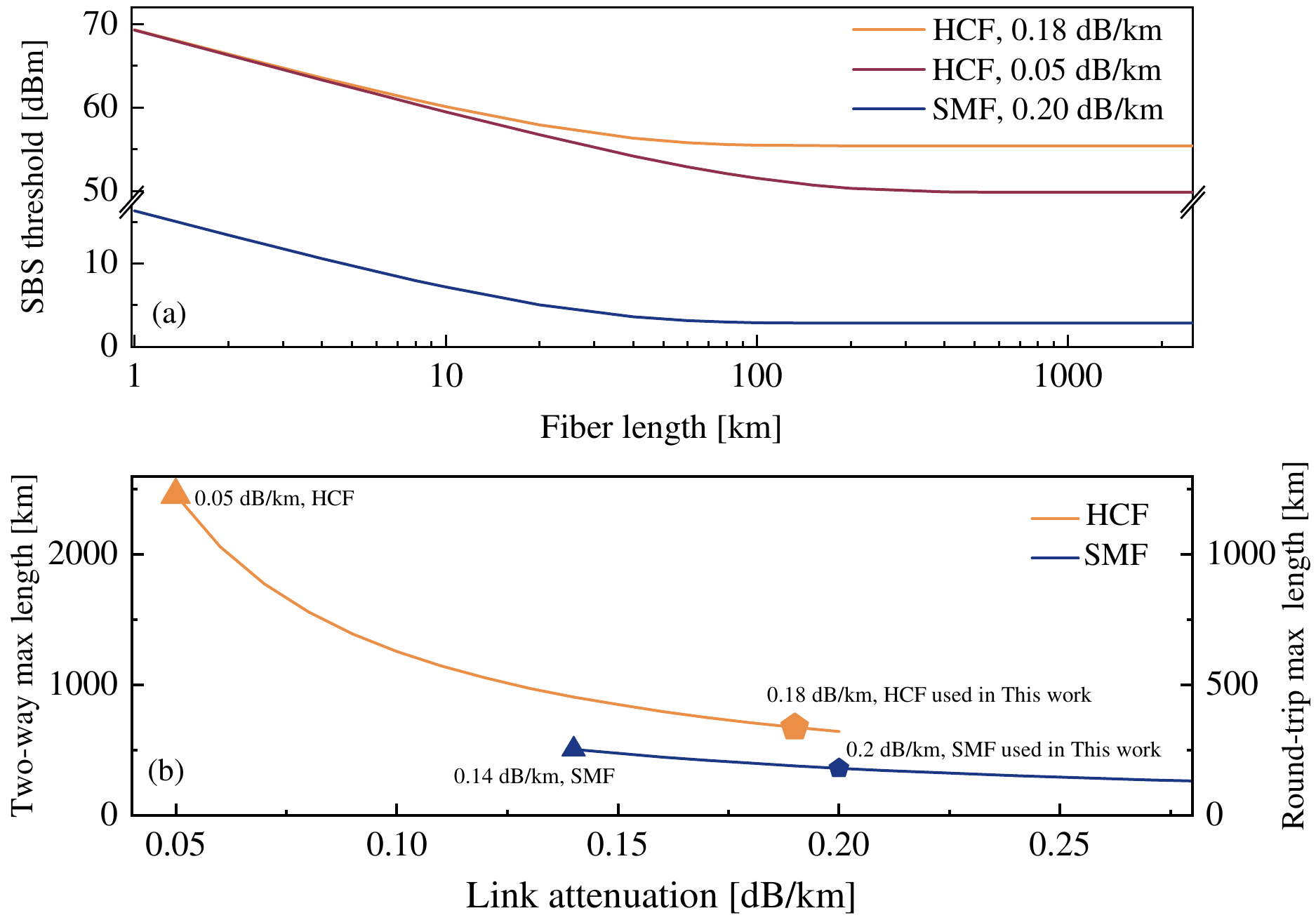}
  \caption{\textbf{a)} Calculated SBS threshold as a function of fiber length for HCF links with attenuations of \(0.18~\mathrm{dB/km}\) and \(0.05~\mathrm{dB/km}\), and for an SMF link with an attenuation of \(0.20~\mathrm{dB/km}\), shown as orange, dark-red, and blue curves, respectively. 
\textbf{b)} Estimated maximum transmission length as a function of link attenuation for HCF and SMF links, shown as orange and blue curves, respectively. 
The left and right axes denote the two-way and round-trip maximum lengths, respectively. }
  \label{fig:boat9}
\end{figure}
Three representative fibers are considered here: conventional SMF with \(\alpha_{\mathrm{dB}}=0.2~\mathrm{dB/km}\), the HCF used in this work with \(\alpha_{\mathrm{dB}}=0.18~\mathrm{dB/km}\), and an advanced ultralow-loss HCF with \(\alpha_{\mathrm{dB}}=0.05~\mathrm{dB/km}\). Figure~9(a) shows the $P_{th,SBS}$ as a function of fiber length $L_{link}$ for these three cases according to eq.~(3). As the fiber length exceeds the effective interaction length \(L_{\mathrm{eff}}\sim 1/\alpha\), the SBS threshold gradually approaches a constant value. In this regime, although the ultralow-loss HCF exhibits an asymptotic SBS threshold about \(5~\mathrm{dB}\) lower than that of the \(0.18~\mathrm{dB/km}\) HCF, the SBS thresholds of HCF still remain overall about \(50~\mathrm{dB}\) higher than that of SMF, preserving a pronounced launch power advantage in the long distance regime.

Figure~9(b) shows the maximum single-span transmission distance of SMF and HCF links for both two way and round trip schemes. Even at the attenuation limit of conventional solid-core silica fiber, about \(0.14~\mathrm{dB/km}\), the maximum single-span distance of SMF is only about \(500~\mathrm{km}\) for the two-way scheme and \(250~\mathrm{km}\) for the round-trip scheme. By contrast, for the HCF used in this work with \(\alpha_{\mathrm{dB}}=0.18~\mathrm{dB/km}\), the corresponding distances increase to about \(700~\mathrm{km}\) and \(350~\mathrm{km}\), respectively. For the ultralow-loss HCF with \(\alpha_{\mathrm{dB}}=0.05~\mathrm{dB/km}\), the transmission reach is expected to extend beyond \(2400~\mathrm{km}\) for the two-way scheme and \(1200~\mathrm{km}\) for the round-trip scheme. These results clearly demonstrate the power-budget advantage of HCF and its strong potential for ultra-long single-span repeater-free optical frequency transfer.

\begin{figure}[!htbp]
  \centering
  \includegraphics[width=0.6\linewidth]{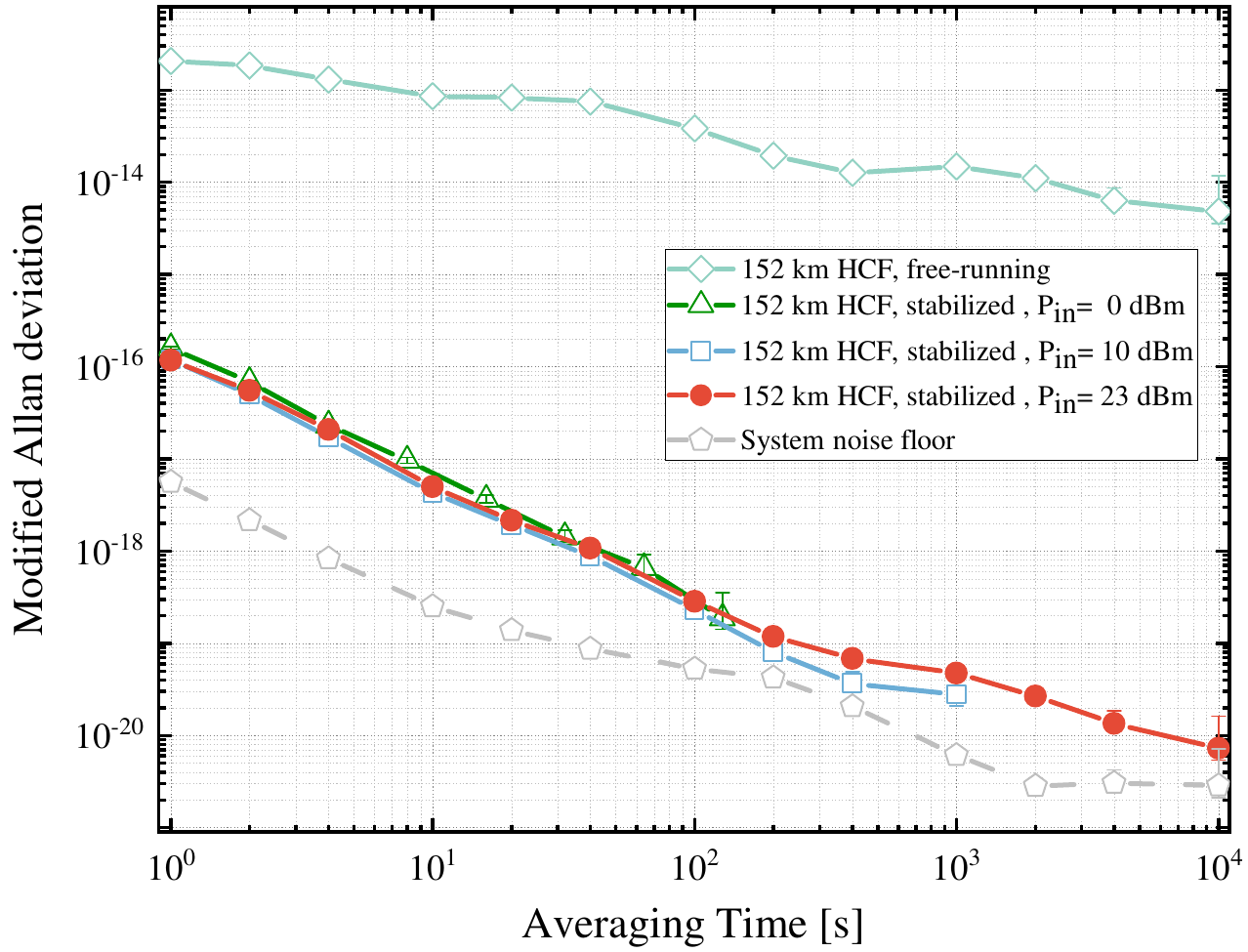}
  \caption{Measured fractional frequency instability of the 152~km HCF link at different input powers. 
Modified Allan deviations are shown for the free-running HCF link, represented by light-green diamonds, and for the stabilized HCF link at input powers of \(P_{\mathrm{in}}=0\), 10, and \(23~\mathrm{dBm}\), represented by green open triangles, blue open squares, and orange filled circles, respectively. 
The gray dashed curve indicates the system noise floor. }
  \label{fig:boat10}
\end{figure}

\section{Demonstration of 152 km single-span optical frequency transfer}

Building on the greatly enlarged power budget demonstrated above, we further evaluated the optical frequency transfer performance of the HCF link under high power operation. In the present experiment, the saturated amplifier output was 34 dBm, corresponding to a maximum injected power of 23 dBm after accounting for the intrinsic loss of the transmission system before the fiber input. The corresponding results are shown in Fig.~10, where the light green diamond symbols represent the transfer instability of the 152~km HCF link under free-running conditions. The instability is approximately \(2.0\times10^{-13}\) at \(1~\mathrm{s}\) and reaches \(4.8\times10^{-15}\) at an averaging time of \(10{,}000~\mathrm{s}\). Compared with previously reported free-running results obtained in laboratory SMF links, the instability of the HCF link continues to decrease at long averaging times, indicating that the low thermal sensitivity of HCF can effectively reduce temperature related fiber induced noise. Furthermore, we evaluated stabilized optical frequency transfer performance at injected optical powers of \(0~\mathrm{dBm}\), \(10~\mathrm{dBm}\), and \(23~\mathrm{dBm}\). The results show that the short-term instability at \(1~\mathrm{s}\) remains at approximately \(1.2\times10^{-16}\) for all tested power levels, with no observable power-dependent degradation. In all cases, the instability follows a consistent \(\tau^{-3/2}\) scaling behavior at short averaging times. In addition, a long-term instability measurement was performed at an injected optical power of \(23~\mathrm{dBm}\), where the transfer instability reached \(7.3\times10^{-21}\) at an averaging time of \(10{,}000~\mathrm{s}\) and gradually approached the system noise floor. These results indicate that HCF can support the transmission of high power optical signals without introducing observable power-dependent excess noise, thereby maintaining stable phase-coherent optical frequency transfer performance under high power operation.

\begin{figure}[!htbp]
  \centering
  \includegraphics[width=0.6\linewidth]{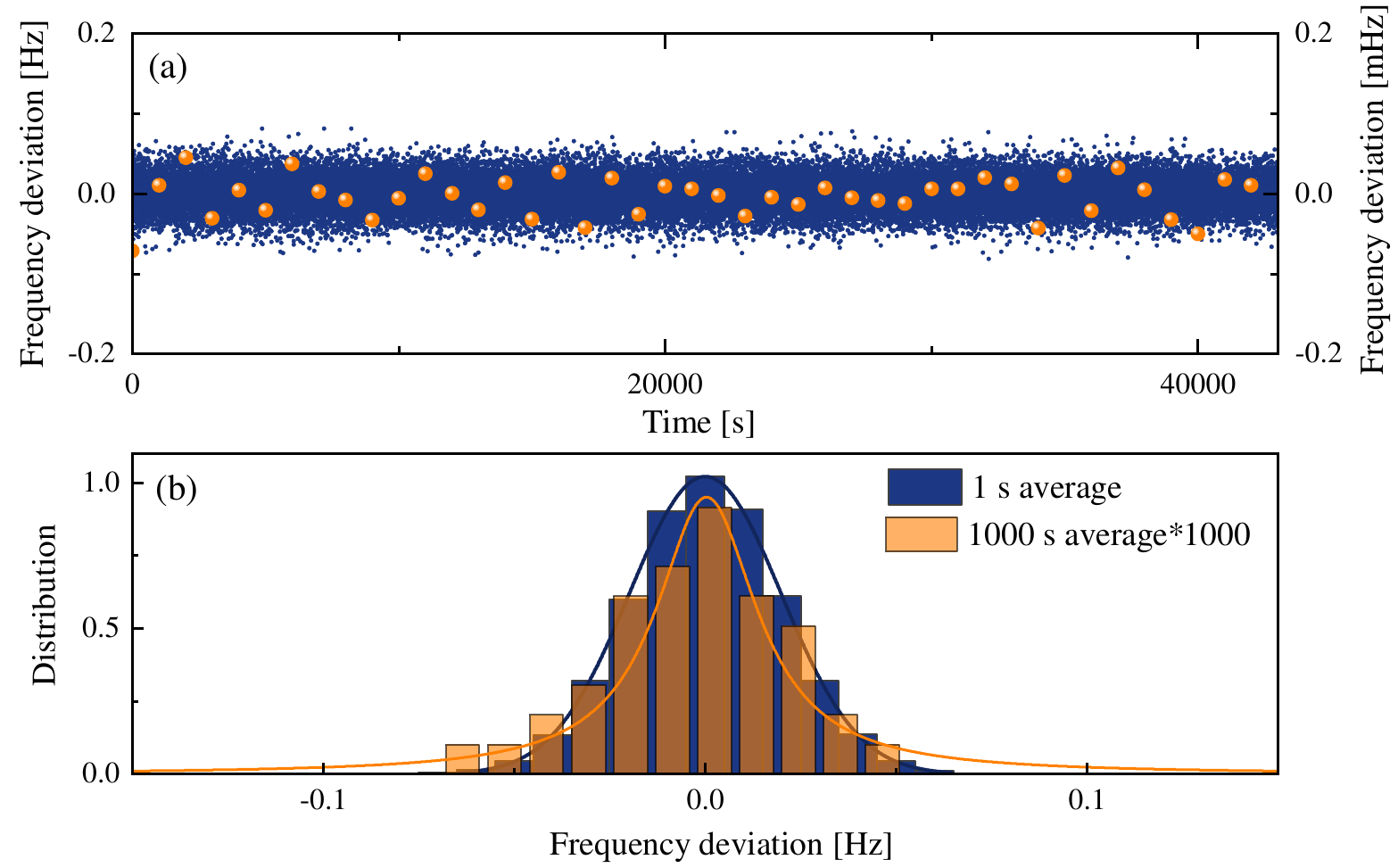}
  \caption{(a) Recorded 43{,}935 frequency data points from the stabilized 152 km HCF link using a dead-time-free $\Lambda$-type frequency counter with a gate time of 1~s (blue points, left axis). The unweighted arithmetic means of all cycle-slip-free 1,000 s segments were calculated, yielding 43 averaged values (orange points, right frequency axis). (b) Histograms and Gaussian fits of the recorded frequency values (blue) and the corresponding 43 averaged values (orange). }
  \label{fig:boat11}
\end{figure}

Complementary to the time domain characterization, we also evaluate the uncertainty of the stabilized 152 km HCF link. Fig.~11(a) shows the measured frequency deviation of the 160~MHz signal for the stabilized HCF link over a continuous duration of \(43{,}935~\mathrm{s}\) (blue dots, left axis). To minimize the statistical uncertainty, all cycle-slip-free 1,000~s averages were calculated (orange dots, right axis)\cite{Benkler2015Metrologia,Lee2010Metrologia}. Fig.~11(b) presents the corresponding histograms and Gaussian fits of the frequency deviations for the full \(43{,}935\) data points(blue) and the 43 averaged data points(orange), respectively. According to the Gaussian fit in Fig.~11(b), the mean frequency offset is \(2.85~\mu\mathrm{Hz}\)(\(1.5\times10^{-20}\)) while the standard deviation of the 1000~s averages is \(23.6~\mu\mathrm{Hz}\)(\(1.2\times10^{-19}\)). The statistical fractional frequency uncertainty is \(3.6~\mu\mathrm{Hz}\)
\(\left(1.8\times10^{-20}\right)\). These results indicate that no measurable systematic frequency shift is introduced by the extraction setup at the level of a few \(10^{-20}\).


\section{Conclusion}
In conclusion, we have experimentally and theoretically demonstrated the advantages of HCF for long-span optical frequency transfer by directly comparing a 152 km HCF link with an equal-length SMF link. To the best of our knowledge, this work represents the first optical frequency transfer experiment over a hundred-kilometer-scale HCF link. The 152 km HCF system achieves a transfer instability of \(7.3\times10^{-21}\) at an averaging time of 10,000 s and reaches an uncertainty at the \(10^{-20}\) level.Compared with SMF, HCF provides nearly one order of magnitude lower thermal sensitivity and effective thermal non-reciprocity, together with reduced fiber-induced phase noise and a shorter propagation delay. These combined advantages lead to a 1.9-fold improvement in short-term frequency instability and an approximately one-order-of-magnitude improvement in long-term frequency instability. At the same time, HCF shows a decisive power-budget advantage, with no observable SBS-induced power saturation up to \(34~\mathrm{dBm}\), an estimated SBS threshold of about \(55.4~\mathrm{dBm}\), and a much larger allowable loss budget than SMF. Consequently, the maximum single span distance is predicted to reach about \(700~\mathrm{km}\) for the present HCF and beyond \(2400~\mathrm{km}\) for advanced ultralow loss HCF. Taken together, these results establish HCF as a highly promising platform for ultra long single-span repeater free optical frequency transfer and future long-haul optical clock networks with improved instability, larger power margin, and lower system complexity.

\medskip
\textbf{Acknowledgements} \par 
This work was funded by the National Natural Science Foundation of China (Grant Nos.~12303077, 62375167 and 62120106010) and the Innovation Program for Quantum Science and Technology
(Grant No.~2021ZD0300900).

\medskip
\textbf{Conflict of Interest} \par 
The authors declare no conflict of interest.

\medskip
\textbf{Data Availability Statement} \par 

The data that support the findings of this study are available from the corresponding author upon reasonable request.
\medskip

%
\bibliographystyle{unsrt}

\bibliography{reference}




\end{document}